\documentclass[showpacs,amsmath,amssymb,11pt]{revtex4}

\usepackage{graphicx}
\usepackage{amssymb}
\usepackage{epstopdf}
\usepackage{epsfig}
\allowdisplaybreaks

\begin{document}

\def\aprge{\buildrel > \over {_{\sim}}}
\def\aprle{\buildrel < \over {_{\sim}}}

\def\etal{{\it et.~al.}}
\def\ie{{\it i.e.}}
\def\eg{{\it e.g.}}

\def\bwt{\begin{widetext}}
\def\ewt{\end{widetext}}
\def\be{\begin{equation}}
\def\ee{\end{equation}}
\def\bea{\begin{eqnarray}}
\def\eea{\end{eqnarray}}
\def\bean{\begin{eqnarray*}}
\def\eean{\end{eqnarray*}}
\def\bary{\begin{array}}
\def\eary{\end{array}}
\def\bi{\bibitem}
\def\bit{\begin{itemize}}
\def\eit{\end{itemize}}

\def\lan{\langle}
\def\ran{\rangle}
\def\lra{\leftrightarrow}
\def\la{\leftarrow}
\def\ra{\rightarrow}
\def\dash{\mbox{-}}
\def\ol{\overline}

\def\ub{\ol{u}}
\def\db{\ol{d}}
\def\sb{\ol{s}}
\def\cb{\ol{c}}

\def\re{\rm Re}
\def\im{\rm Im}

\def \b{{\cal B}}
\def \ca{{\cal A}}
\def \ko{K^0}
\def \ok{\overline{K}^0}
\def \s{\sqrt{2}}
\def \st{\sqrt{3}}
\def \sx{\sqrt{6}}
\title{{\bf Strong and Electromagnetic Decays of X(1835) as a Baryonium State}}
\author{Yong-Liang Ma}
\address{
Institute for Theoretical Physics, University of Tuebingen,
   D-72076 Tuebingen, Germany}
\date{\today}
\begin{abstract}
With the assumption that the recently observed X(1835) is a
baryonium state we have studied the strong decays of $X(1835) \to
\eta^{(\prime)} \pi^+ \pi^-, \eta^{(\prime)} \pi^0 \pi^0$ and the
electromagnetic decay of $X(1835) \to 2\gamma$ in the framework of
effective Lagrangian formalism. In the present investigation we have
included the contributions from the iso-singlet light scalar
resonances but we have not included the isospin violating effect.
Our result for the strong decay of $X(1835) \to \eta^{\prime} \pi^+
\pi^-$ is smaller than the observed data. The decay width for the
radiative decay of $X(1835) \to 2\gamma$ is consistent with the
assumption that it decays through the glueball. In addition, the
width for the strong decay of $X(1835) \to \eta \pi^+ \pi^-$ is
larger than that of the strong decay of $X(1835) \to \eta^{\prime}
\pi^+ \pi^-$ due to the large phase space and coupling constant $g_{
N\bar{N}\eta}$. From our investigation, it is not possible to
interpret X(1835) as a baryonium.
\end{abstract}
\maketitle

\section{Introduction}

In 2005, the BES collaboration announced the observation of a
resonant state termed X(1835) in the reaction $J/\psi \to \gamma X$,
$X \to \eta^\prime\pi^+\pi^-$~\cite{Ablikim:2005um}. A fit to this
resonance with the Breit-Wigner function yields the quantum number
$J^{PC} = 0^{-+}$ and mass $M_X = (1833.7 \pm 6.1 \pm 2.7)~$MeV,
width $\Gamma_X = (67.7 \pm 20.3 \pm 7.7)~$MeV and the product
branching fraction ${\rm Br}(J/\psi \to \gamma X(1835)){\rm
Br}(X(1835) \to \eta^\prime\pi^+\pi^-) = (2.2 \pm 0.4({\rm stat.})
\pm 0.4({\rm syst.}))\times 10^{-4}$. Actually, without include the
final state interaction, the parameters of this resonance have been
fitted to be $M_X \simeq 1859^{+3}_{-10}({\rm stat})^{+5}_{-25}({\rm
syst})~{\rm MeV}$ and the total width  $\Gamma < 30~{\rm MeV}$ in
Ref.~\cite{Bai:2003sw}. Since the discovery of the X(1835) state,
many models have been proposed to explain its
properties~\cite{Rosner:2005gf,Rosner:2003bm,Datta:2003iy,Zou:2003zn,
Liu:2004er,Chang:2004us,Sibirtsev:2004id,Yan:2005wr,Ding:2005ew,Zhu:2005ns,Wang:2006sna,
Kochelev:2005vd,Li:2005vd,He:2005nm,Huang:2005bc,Klempt:2007cp,Li:2008mz}.

In the previous works, the X(1835) state has been conjectured to be
a baryonium
state~\cite{Rosner:2005gf,Rosner:2003bm,Datta:2003iy,Zou:2003zn,
Liu:2004er,Chang:2004us,Sibirtsev:2004id,Yan:2005wr,Ding:2005ew,Zhu:2005ns,Wang:2006sna},
pseudoscalar glueball
state~\cite{Kochelev:2005vd,Li:2005vd,He:2005nm} and also a radial
excitation of
$\eta^\prime$~\cite{Huang:2005bc,Klempt:2007cp,Li:2008mz}. Although
there are many speculations, none of the above claims can be either
confirmed or ruled out by the present experiments. In our present
work, we have calculated the strong decays of $X(1835) \to
\eta^{(\prime)} \pi^+\pi^-, \eta^{(\prime)} \pi^0\pi^0$ and
radiative decay of $X(1835) \to 2\gamma$ using the effective
Lagrangian formalism by treating the X(1835) as a $N\bar{N}$
baryonium. This seems to be a reasonable approximation if one only
considers the fact that the mass of the X(1835) is bit lower than
the threshold energy of $p\bar{p}$ and $n\bar{n}$ (about 40 MeV).
Our philosophy is that, assuming the X(1835) as a baryonium, if we
can get the numerical results agree with the observed data the
baryonium assumption is reasonable otherwise the baryonium picture
can be ruled out, at least in this framework. The coupling of the
X(1835) to its constituents can be described by the effective
Lagrangian. The corresponding effective coupling constant $g_{_{X}}$
is determined by the compositeness condition $Z = 0$ which was
earlier used by nuclear physicists
~\cite{Weinberg:1962hj,Salam:1962ap,Hayashi:1967hk} and is being
widely used by particle physicists (see the references in
~\cite{Faessler:2007gv}). We had applied the above method to study
the newly observed charmed mesons
~\cite{Faessler:2007gv,Faessler:2007us,Dong:2008gb} and their decay
properties which we had obtained agreed with the observed data. We
had also employed the above technique to predict the decay
properties of the bottom-strange mesons~\cite{Faessler:2008vc}. In
our present work, we have used a typical scale parameter $\Lambda_X$
to describe the finite size of the baryonium. The value of
$\Lambda_X$ is fixed by considering the coupling constant $g_{_{X}}$
is expected to be stable. For other interactions, we have used the
phenomenological Lagrangian and have borrowed the relevant coupling
constants from the existing literature. Using the above
phenomenological approaches, we have analyzed the strong decays of
$X(1835) \to \eta^{(\prime)} \pi^+\pi^-, \eta^{(\prime)} \pi^0\pi^0$
and radiative decay of $X(1835) \to 2\gamma$. The result of the
decay width of $X(1835) \to \eta^{\prime}\pi^+\pi^-$ is much smaller
than the observed data hence the X(1835) cannot be treated as a
baryonium.

The paper is organized in the following way: In Section
\ref{TheoFram}, we have calculated the effective coupling constant
$g_{_{X}}$ using the compositeness condition and have discussed the
effective Lagrangian formalism employed in our calculation. In
Section \ref{StrongEM} we have calculated the strong decay widths of
$X(1835) \to \eta^{(\prime)}\pi^+\pi^-$ and radiative decay width of
$X(1835) \to 2\gamma$ using the effective coupling constant
$g_{_{X}}$ and effective Lagrangian proposed in Section
\ref{TheoFram}. In section \ref{DisCon} the important results and
conclusions have been given.

\section{Theoretical framework}

\label{TheoFram}

\subsection{Baryonium structure of the X(1835) state}

In this section we give the formulation for the study of the X(1835)
as a baryonium state which can be thought of as a
$p\bar{p}(n\bar{n})$ bound state. As stated earlier, the mass of the
X(1835) is around $40~$MeV less than the threshold of
$p\bar{p}(n\bar{n})$. The quantum number of the X(1835) is assigned
to be $J^{PC} = 0^{-+}$, and its mass is predicted to be $m_{X} =
1833.7$ MeV~\cite{Ablikim:2005um}. The effective Lagrangian
describing the interaction between the X(1835) and its constituents
is given by
\begin{eqnarray}
{\cal L}_{X(1835)}(x) & = & ig_{_{X}}X(x)\int
dy\Phi_X(y^2)\bar{N}(x+\frac{1}{2}y)\gamma_5N(x-\frac{1}{2}y)\,,
\label{LeffXNN}
\end{eqnarray}
where the baryon doublet $N$ is defined as
\begin{eqnarray}
N = \left(
         \begin{array}{c}
           p \\
           n \\
         \end{array}
       \right) \nonumber
\end{eqnarray}
The correlation function $\Phi_X$ characterizes the finite size of
the X(1835) as a $N\bar{N}$ bound state and depends on the relative
Jacobi coordinates $y$ and $x$. In the limit $\Phi_X(y^2) \to
\delta^4(y)$, the interaction given by Eq.~(\ref{LeffXNN}) becomes
local. The Fourier transform of the correlation function
$\Phi_X(y^2)$ is
\begin{eqnarray}
\Phi_X(y^2) & = & \int\frac{d^4p}{(2\pi)^4}e^{-ip\cdot
y}\tilde{\Phi}_X(-p^2) \nonumber
\end{eqnarray}
In following calculation, an explicit form of $\tilde{\Phi}_X$ has
been used. The choice of $\tilde{\Phi}_X$ should be such that it
falls off sufficiently fast in the ultraviolet region of Euclidean
space to render the Feynman diagrams finite in the UV region. In
this sense, one can also regard $\tilde{\Phi}_X$ as a regulator for
the loop integral. In our work, we have chosen the Gaussian form for
$\tilde{\Phi}_X$
\begin{eqnarray}
\tilde{\Phi}_X(p_E^2) = \exp(-p_E^2/\Lambda_X^2) \nonumber
\end{eqnarray}
where $p_E$ is the Euclidean Jacobi momentum. Here $\Lambda_X$ is a
size parameter which parameterizes the distribution of $N$ and
$\bar{N}$ baryons inside the X(1835) baryonium.

The coupling constant $g_{{_X}}$ is determined by the compositeness
condition~\cite{Weinberg:1962hj,Salam:1962ap,Hayashi:1967hk} which
implies that the renormalization constant of the hadron wave
function is set to zero
\begin{eqnarray}
Z_{X} & = & 1 - \Sigma^\prime_{X}(m_{X}^2) = 0 \,.\label{ZX}
\end{eqnarray}
Here, $\Sigma^\prime_{X}(m_X^2) = g_{_{X}}^2
\Pi^\prime_{X}(m_{X}^2)$ is the derivative of the mass operator
$\Sigma_{X}$ which is represented by the diagrams in
Fig.~\ref{fig:massOperator} given below.
\begin{figure}[htbp]
\begin{center}
\includegraphics[scale=0.7]{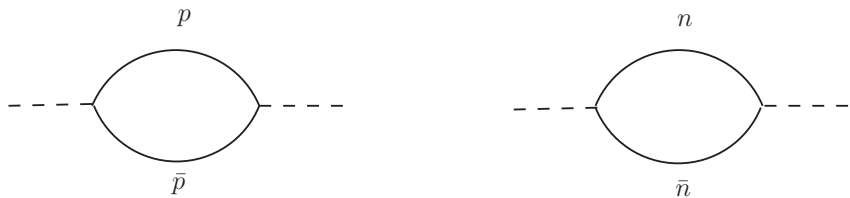}
\end{center}
\caption[Mass operator of X(1835) as a baryonium state.]{%
Mass operator of X(1835) as a baryonium state. }
\label{fig:massOperator}
\end{figure}

The compositeness condition can be understood in the following: The
renormalization constant $Z_X^{1/2}$ can be interpreted as the
matrix element between the physical state X(1835) and corresponding
bare state $X_0(1835)$, i.e., $\langle 0| X_0(1835)|X(1835)\rangle =
Z_X^{1/2}\langle 0| X(1835)|X(1835)\rangle = Z_X^{1/2}$, so that
$Z_X = 0$ means that the physical state should not be a function of
the corresponding bare state which means that the physical state is
a bound state. In our present work, the X(1835) is a bound state of
$p\bar{p}(n\bar{n})$. In this sense, the compositeness condition
excludes the possibility of the processes involving the X(1835) as
an initial or a final state since each external X(1835) contributes
a factor $Z^{1/2}_{X}$ to the relevant matrix elements. In addition,
because of the interaction between the X(1835) and its constituents,
the mass and wave function of the X(1835) have to be renormalized.

\begin{figure}[htbp]
\begin{center}
\includegraphics[scale=0.5]{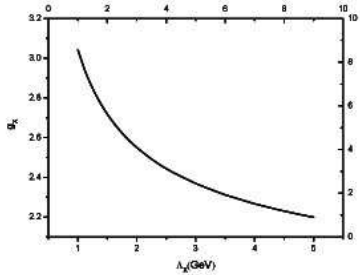}
\end{center}
\caption[The $\Lambda_X$ dependence of the coupling constant $g_{_X}$.]{%
The $\Lambda_X$ dependence of the coupling constant $g_{_X}$. }
\label{fig:gxlambdax}
\end{figure}

Following Eq.~(\ref{ZX}) the coupling constant $g_{_{X}}$ can be
expressed as
\begin{eqnarray}
\frac{1}{g_{_{X}}^2} & = & \frac{1}{4 \pi^2}\int_0^\infty
d\alpha\int_0^1dx\frac{1}{(1+\alpha)^2}\frac{d}{d\mu_X^2}\bigg\{\tilde{\Phi}(z_1)+\alpha\mu_X^2\tilde{\Phi}(z_2)\bigg\}\label{gX_coupling}
\end{eqnarray}
where $\alpha$ and $x$ are both Feynman parameters and
\begin{eqnarray}
z_1 & = & \alpha m_p^2 - \frac{\alpha}{4(1+\alpha)}m_X^2\nonumber\\
z_2 & = & \alpha m_p^2 - \frac{\alpha +
4\alpha^2x(1-x)}{4(1+\alpha)}m_X^2\nonumber\\
\mu_X^2 & = & m_X^2/\Lambda_X^2  \nonumber
\end{eqnarray}
$x$ and $\alpha$ are both Feynman parameters. In deriving the
expression (\ref{gX_coupling}), we have ignored the mass difference
between proton and neutron and expressed the coupling constant
$g_{_{X}}$ in terms of the proton mass. To get the numerical result
of $g_{_{X}}$, we use $m_X = 1833.7~$MeV~\cite{Ablikim:2005um}, $m_p
= 938.272~$MeV~\cite{Yao:2006px} and vary the scale parameter
$\Lambda_X$ from $1.0~$GeV to $5.0~$GeV. In Fig.~\ref{fig:gxlambdax}
we show the $\Lambda_X$ dependence of the effective coupling
constant $g_{_X}$.

Concerning that $g_{_{X}}$ is expected to be stable against
$\Lambda_X$, we choose the region of $\Lambda_X$ as $2.0 {\rm ~GeV}
\leq \Lambda_X \leq 3.0{\rm ~GeV}$ and get the coupling constant to
be in the range $g_{_{X}} = 2.37 - 2.55$. Comparing our present
result with that given in Ref.~\cite{Zhu:2005ns} where this coupling
constant was estimated from experimental branching ratio of the
X(1835) to $p\bar{p}$ decay in radiative decay of $J/\psi$ (by
considering that $X \to p\bar{p}$ occurs from the tail of its mass
distribution and the value was found to be, $|g_{_{Xp\bar{p}}}|
\simeq 3.5$), we conclude our result agrees with the result given
there. In fact, using ${\rm BR}(X \to p\bar{p}) \sim (0.04-0.14)$
assuming $\Gamma_X < 30 $MeV\cite{Bai:2003sw} that
Ref.~~\cite{Zhu:2005ns} adopted, one can get $g_{_{Xp\bar{p}}} = 2.2
- 4.1$. In addition, our conclusion is also consistent with that of
Ref.~\cite{Li:2005vd} which was based on the glueball assumption.
Expressing the coupling constant $g_{_{Xp\bar{p}}}$ in terms of
$g_{_{X_g}}$ which is the coupling constant between the X(1835) and
glueball, one can get $g_{_{Xp\bar{p}}} = 2.47 - 4.67$.

\subsection{Effective Lagrangian for strong and electromagnetic decays of X(1835)}

In this section, we have discussed the effective Lagrangian for the
calculation of the strong decays of $X(1835) \to \eta^{(\prime)}
\pi^+\pi^-$ and electromagnetic decay of $X(1835) \to 2\gamma$. The
effective lagrangian can be divided into two parts, the free part
${\cal L}_{\rm free}$ and the interaction part ${\cal L}_{\rm int}$.
It should be noted that the electromagnetic interaction can be
obtained by the minimal substitution (i.e., replacing the derivative
operator $\partial_\mu$ of the charged particle with the covariant
one $D_\mu = \partial_\mu - ieQA_\mu$ with $Q$ as the charge of the
relevant particle). For the free Lagrangian, it involves states with
quantum numbers $J^P = \frac{1}{2}^+, 0^-, 0^+$ and $1^-$.
\begin{eqnarray}
{\cal L}_{\rm free} = {\cal L}_{\rm free}^N + {\cal L}_{\rm free}^P
+ {\cal L}_{\rm free}^S + {\cal L}_{\rm free}^V  \nonumber
\end{eqnarray}
where
\begin{eqnarray}
{\cal L}_{\rm free}^N & = &
\bar{N}(i\partial\hspace{-0.2cm}\slash-m_N)N  \nonumber\\
{\cal L}_{\rm free}^P & = & -\frac{1}{2}X(1835)(\Box +
m_X^2)X(1835)\nonumber\\
& & - \frac{1}{2}\vec{\pi}(x)(\Box + m_\pi^2)\vec{\pi}(x) -
\frac{1}{2}\eta(x)(\Box + m_{\eta^\prime}^2)\eta(x) -
\frac{1}{2}\eta^\prime(x)(\Box + m_{\eta^\prime}^2)\eta^\prime(x) \nonumber\\
{\cal L}_{\rm free}^S & = & -\frac{1}{2}\sigma(\Box +
m_\sigma^2)\sigma -\frac{1}{2}f_0(\Box +
m_{f_0}^2)f_0  \nonumber\\
{\cal L}_{\rm free}^V & = & -\frac{1}{4}F_{\mu\nu}F_{\mu\nu}
\nonumber
\end{eqnarray}
with $F_{\mu\nu} = \partial_\mu A_\nu - \partial_\nu A_\mu$ as the
field tensor of photon and $\Box \equiv \partial_\mu\partial^\mu$.
For computing the decays of the X(1835), we have treated the masses
of proton and neutron and the masses of the triplet pions to be the
same ~\cite{Ablikim:2005um,Yao:2006px}
\begin{eqnarray}
m_X & = & 1833.7 \mbox{~MeV}; \;\;\;\;\;\;\;\;\;\;\;\;\;\;\;\;\;\;
m_n = m_p = 938.27203
\mbox{~MeV}\nonumber\\
m_{\pi^0} & = & m_\pi^\pm = 139.57018 \mbox{~MeV}; \;\; m_{\eta} =
547.51 \mbox{~MeV}; \;\; m_{\eta^\prime} = 957.78
\mbox{~MeV};\label{spectrum}
\end{eqnarray}
while for the masses and widths of scalar mesons, we have
adopted~\cite{AbdelRehim:2002an}
\begin{eqnarray}
m_\sigma & = & 550 \mbox{~MeV}; \;\;\;\;\;\;\Gamma_{\sigma} = 370
\mbox{~MeV}; \;\;\;\;\;\;\;\; m_{f_0} = 980 \mbox{~MeV};\;\;\;\;\;\;
\Gamma_{f_0} = 64.6 \mbox{~MeV}  \nonumber
\end{eqnarray}

In the following calculation, we have included the finite width
effects of the scalar mesons, that is, we have written the scalar
meson propagators as
\begin{eqnarray}
\tilde{D}_S(k) & = & \frac{i}{k^2-m_S^2+im_S\Gamma_S}  
\nonumber
\end{eqnarray}

The interaction Lagrangian ${\cal L}_{\rm int}$ used in our
calculation has two parts, the strong part ${\cal L}_{\rm int}^{\rm
str}$ and the electromagnetic part ${\cal L}_{\rm int}^{\rm em}$
\begin{eqnarray}
{\cal L}_{\rm int} & = & {\cal L}_{\rm int}^{\rm str} + {\cal
L}_{\rm int}^{\rm em} \nonumber
\end{eqnarray}
For the strong interaction Lagrangian we have ${\cal L}_{N\bar{N}X}$
(X-nucleon-nucleon interaction), ${\cal L}_{N\bar{N}P}$
(pseudoscalar-nucleon-nucleon interaction), ${\cal L}_{N\bar{N}S}$
(scalar-nucleon-nucleon interaction) and ${\cal L}_{SPP}$
(scalar-pseudoscalar-pseudoscalar interaction)
\begin{eqnarray}
{\cal L}_{\rm int}^{\rm str} & = & {\cal L}_{N\bar{N}X} + {\cal L}_{
N\bar{N}P} + {\cal L}_{ N\bar{N}S} + {\cal L}_{SPP} 
\nonumber
\end{eqnarray}
The effective Lagrangian ${\cal L}_{N\bar{N}X}$ was given in
Eq.~(\ref{LeffXNN}) and ${\cal L}_{N\bar{N}P}$ and ${\cal
L}_{N\bar{N}S}$ can be expressed as
\begin{eqnarray}
{\cal L}_{N\bar{N}P} & = & \frac{1}{2m}g_{N\bar{N}\pi}\bar{N}
\gamma_\mu\gamma_5\vec{\tau} N\partial_\mu \vec{\pi} +
\frac{1}{2m}g_{N\bar{N}\eta}\bar{N} \gamma_\mu\gamma_5 N
\partial_\mu \eta + \frac{1}{2m}g_{N\bar{N}\eta^\prime}\bar{N}
\gamma_\mu\gamma_5 N \partial_\mu \eta^\prime \label{LpiN}\\
{\cal L}_{N\bar{N}S} & = & g_{N\bar{N}S}\bar{N} NS \label{LSN}\\
{\cal L}_{SPP} & = & -\frac{\gamma_{\sigma\pi\pi}}{\sqrt{2}}\sigma
\partial_\mu \vec{\pi}\cdot \partial_\mu \vec{\pi}
-\frac{\gamma_{f_0\pi\pi}}{\sqrt{2}}f_0
\partial_\mu \vec{\pi}\cdot \partial_\mu \vec{\pi}
\end{eqnarray}
where $S$ is the scalar meson ($\sigma$ and $f_0$ in our problem)
and $\pi$ is the pseudoscalar meson matrix
\begin{eqnarray}
\pi & = & \sum_{i=1}^3\pi^i\tau^i=\left(
                           \begin{array}{cc}
                             \pi^0 & \sqrt{2}\pi^+ \\
                             \sqrt{2}\pi^- & -\pi^0 \\
                           \end{array}
                         \right); \;\; \nonumber
\end{eqnarray}
The coupling constants $g_{N\bar{N}\pi}$, $g_{N\bar{N}\eta}$ and
$g_{N\bar{N}\eta^\prime}$ were determined via the $J/\psi$ hadronic
decay~\cite{Sinha:1984qn,Liang:2004sd} while $g_{N\bar{N}S}$ was
yielded by fitting the theoretical results of $NN$ scattering with
the observed data~\cite{Machleidt:1987hj}
\begin{eqnarray}
(g_{N\bar{N}\pi})^2/(4\pi) & \simeq & 14.8  \nonumber\\
(g_{N\bar{N}\eta }/g_{N\bar{N}\pi })^2 & \simeq & 3.90625 \times 10^{-3}  \nonumber\\
(g_{N\bar{N}\eta^\prime }/g_{N\bar{N}\pi })^2 & \simeq & 2.5 \times
10^{-3} \nonumber\\
(g_{N\bar{N}S })^2/(4\pi) & \simeq & 5.69 \nonumber
\end{eqnarray}
The scalar-pseudoscalar-coupling constant $\gamma_{SPP}$ was given
in Ref.~\cite{AbdelRehim:2002an}
\begin{eqnarray}
\gamma_{\sigma\pi\pi} = 7.27 \mbox{~GeV}^{-1} ; \;\;\;\;\;\;\;\;
\gamma_{f_0\pi\pi} = 1.47 \mbox{~GeV}^{-1} \nonumber
\end{eqnarray}

For the electromagnetic interaction Lagrangian ${\cal L}_{\rm
int}^{\rm em}$ used in our calculation, it has two parts: (i) is
from the gauge of the charged free nucleon Lagrangian, and (ii) is
from the gauge of the nonlocal interaction
\begin{eqnarray}
{\cal L}_{\rm int}^{\rm em} & = & {\cal L}_{\rm int}^{\rm em(i)} +
{\cal L}_{\rm int}^{\rm em(ii)}  \nonumber
\end{eqnarray}
where
\begin{eqnarray}
{\cal L}_{\rm int}^{\rm em(i)} & = & eA_\mu\bar{N} \gamma_\mu
\frac{1+\tau_3}{2}N \label{GaugeProton}\\
{\cal L}_{\rm int}^{\rm em(ii)} & = & ig_{_{X}}X(x)\int
dy\Phi_X(y^2)\big\{e^{ieI(x+\frac{1}{2}y,x-\frac{1}{2}y;P)}\bar{p}(x+\frac{1}{2}y)\gamma_5
p(x-\frac{1}{2}y)\big\}\label{EMXNN}
\end{eqnarray}
where the Wilson line $I(x,y,P)$ is defined as
\begin{eqnarray}
I(x,y;P) & = & \int_y^x dz_\mu A^\mu(z) \nonumber
\end{eqnarray}
To derive the Feynman rules for photons, we require the derivative
of $I(x,y;P)$. For this we have used the path-independent
prescription as suggested in
Ref.~\cite{Mandelstam:1962us,Terning:1991yt} which implies that the
derivative of $I(x,y;P)$ does not depend on the path $P$ originally
used in the definition. Also in our calculation of $X(1835) \to
2\gamma$, in principle we should expand the above expression to the
second order but the diagram with photons from this vertex does not
contribute since the trace of gamma matrices vanish.

\section{Strong and electromagnetic decays}

\label{StrongEM}

Having discussed the effective coupling constant $g_{_{X}}$ and the
effective Lagrangian, we are in the position to calculate the decay
properties of  the X(1835). In this section, we have calculated the
strong decays of $X(1835)\to \eta^{(\prime)} \pi^+\pi^-$  and also
the radiative decay of $X(1835)\to2\gamma$.

\subsection{Strong decays of $X(1835) \to \eta^{(\prime)} \pi^+\pi^-$}

\begin{figure}[htbp]
\begin{center}
\includegraphics[scale=0.6]{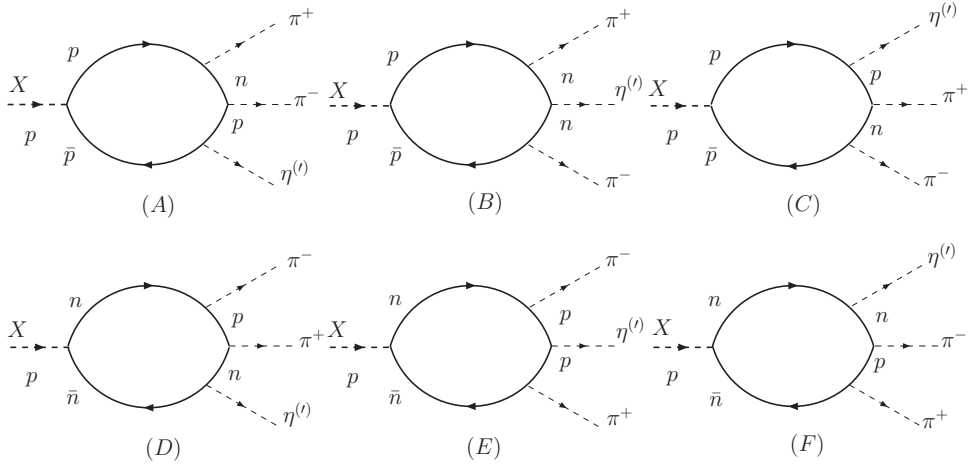}
\end{center}
\caption[Diagrams contribute to the strong decay $X(1835) \to \eta^{(\prime)} \pi^+\pi^-$ without scalar resonance contribution.]{%
Diagrams contributing  to the strong decay of decay of $X(1835) \to
\eta^{(\prime)} \pi^+\pi^-$ without scalar resonance contribution. }
\label{fig:strongdecayp}
\end{figure}

\begin{figure}[htbp]
\begin{center}
\includegraphics[scale=0.55]{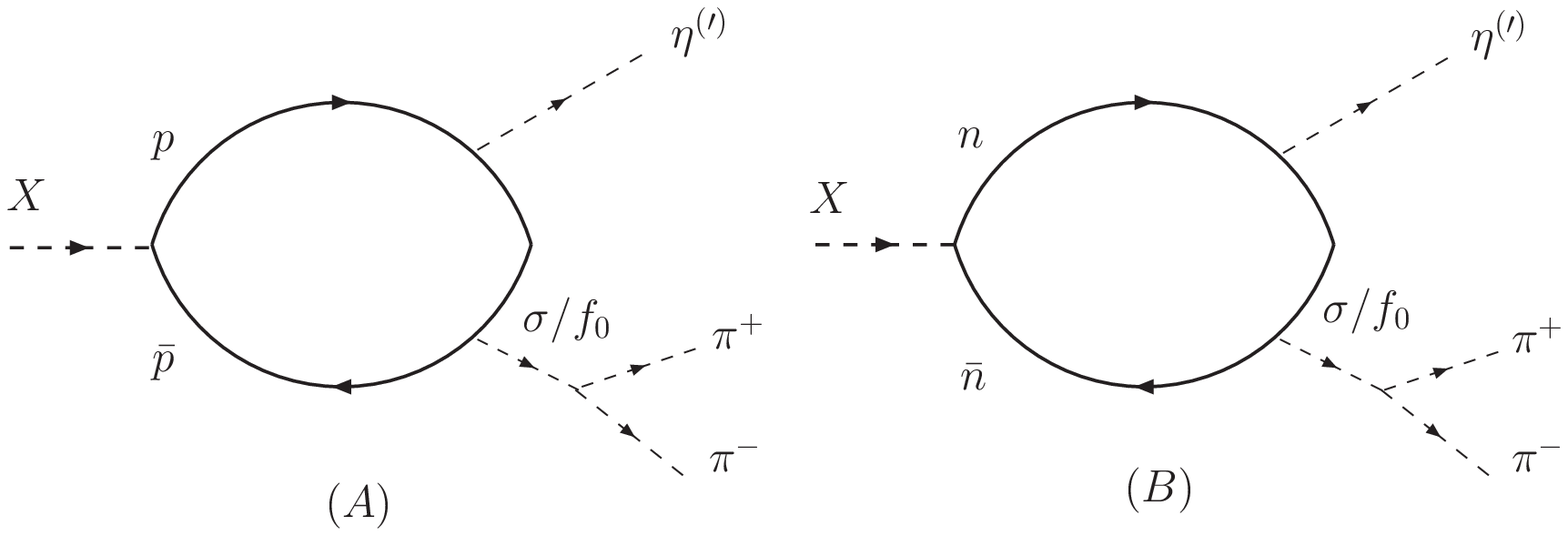} \\
\includegraphics[scale=0.55]{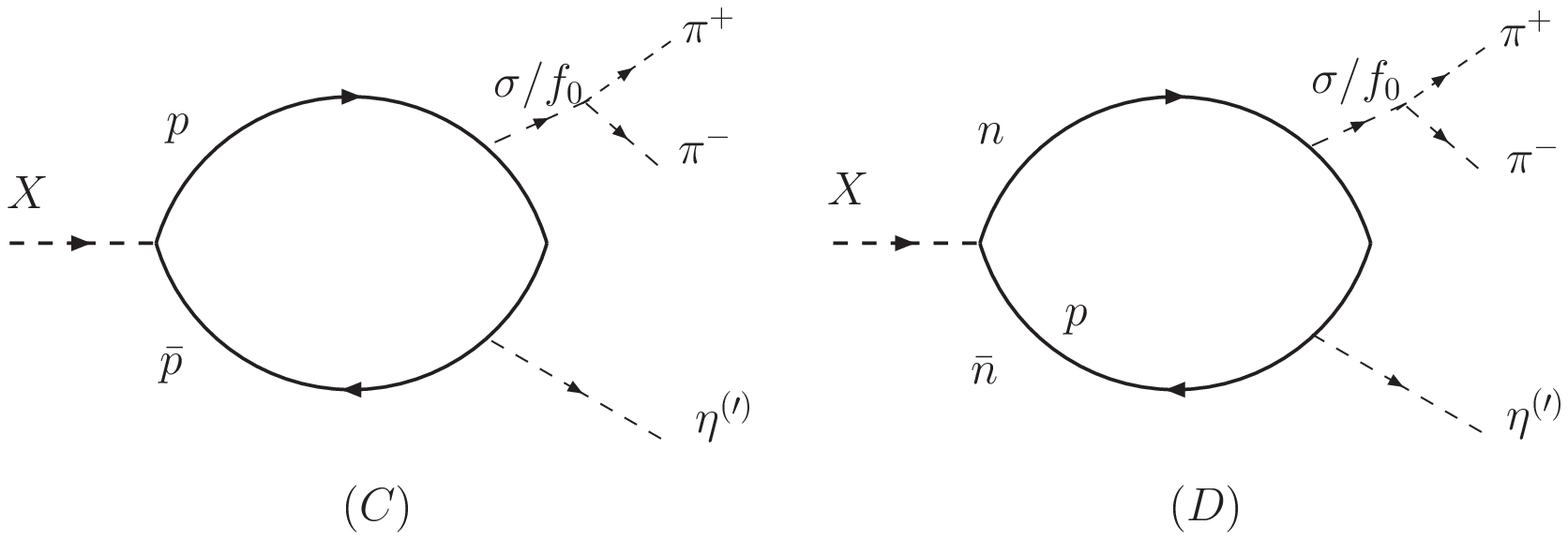}
\end{center}
\caption[Diagrams contribute to the strong decay $X(1835) \to \eta^{(\prime)} \pi^+\pi^-$ with scalar resonance contribution.]{%
Diagrams contributing  to the strong decay of decay of $X(1835) \to
\eta^{(\prime)} \pi^+\pi^-$ with scalar resonance contribution. }
\label{fig:strongdecays}
\end{figure}

For the strong decays of $X(1835) \to \eta^{(\prime)} \pi^+\pi^-$,
the Feynman diagrams of Fig.~\ref{fig:strongdecayp} and
Fig.~\ref{fig:strongdecays} contribute. All the diagrams listed in
Fig.~\ref{fig:strongdecayp} are from the one-pseudoscalar
meson-nucleon-nucleon vertex while the diagrams listed in
Fig.~\ref{fig:strongdecays} are from the scalar resonance
contributions. For the isospin symmetric case following relations
among matrix elements exist
\begin{eqnarray}
iM_{(A)} & = & iM_{(D)}; \;\;\;iM_{(B)} = iM_{(E)}; \;\;\; iM_{(C)}
= iM_{(F)} \nonumber
\end{eqnarray}

\begin{figure}[htbp]
\begin{center}
\includegraphics[scale=0.5]{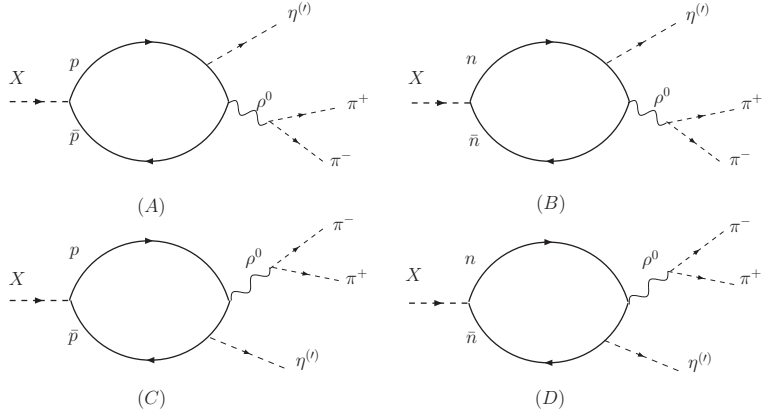}
\end{center}
\caption[Diagrams contributing to the strong decay $X(1835) \to \eta^{(\prime)} \pi^+\pi^-$ from the $\rho$ meson exchange.]{%
Diagrams contributing to the strong decay $X(1835) \to
\eta^{(\prime)} \pi^+\pi^-$ from the $\rho$ meson exchange. }
\label{fig:strongdecayisorho}
\end{figure}

\begin{figure}[htbp]
\begin{center}
\includegraphics[scale=0.5]{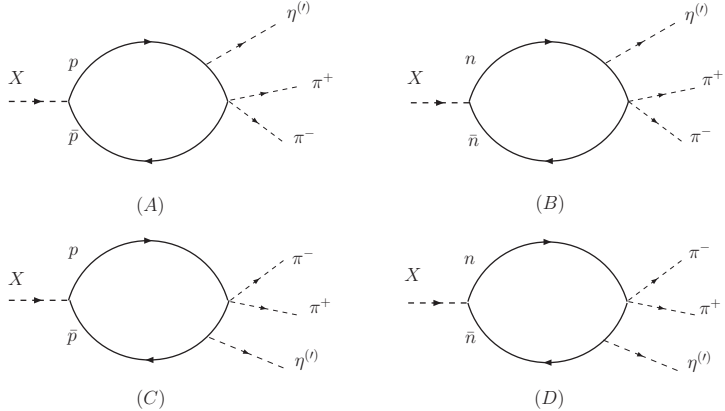}
\end{center}
\caption[Diagrams contributing  to the strong decay of
$X(1835) \to \eta^{(\prime)} \pi^+\pi^-$ from two-meson-nucleon vertex.]{%
Diagrams contributing to the strong decay $X(1835) \to
\eta^{(\prime)} \pi^+\pi^-$ from the two-meson-nucleon vertex. }
\label{fig:strongdecayiso2pi}
\end{figure}

It should be noted that to include the isospin violating effect, the
diagrams in Fig.~\ref{fig:strongdecayisorho} and
Fig.~\ref{fig:strongdecayiso2pi} should also be considered. For
isospin symmetric case the matrix elements for the diagrams of
Fig.~\ref{fig:strongdecayisorho} and
Fig.~\ref{fig:strongdecayiso2pi} have the following relations
\begin{eqnarray}
iM_{(A)} & = & - iM_{(B)}; \;\;\;iM_{(C)} = - iM_{(D)} \nonumber
\end{eqnarray}
In our present work we have considered isospin symmetric case and
hence diagrams of Fig.~\ref{fig:strongdecayisorho} and
Fig.~\ref{fig:strongdecayiso2pi} do not contribute. In addition, the
diagrams with
$\rho_\mu^\mp\eta^{(\prime)}\partial^{^{^{\hspace{-0.2cm}\leftrightarrow}}}_\mu\pi^\pm$
vertex also have not been considered due to the G-parity
conservation.


In the following calculation, we label the momenta of the relevant
particles according to the scheme $X(p) \to
\pi^+(q_1)+\pi^-(q_2)+\eta^{(\prime)}(q_3)$. The partial decay width
is related to the invariant matrix element $M(p \to q_1+q_2+q_3)$ by
the relation
\begin{eqnarray}
\Gamma(X(1835) \to \pi^+\pi^-\eta^{(\prime)}) & = &
\frac{1}{2m_X}\int|M^2|d\Phi \nonumber
\end{eqnarray}
where $d\Phi$ is the Lorentz invariant phase space volume element
\begin{eqnarray}
d\Phi & = &
(2\pi)^4\delta^4(p-\sum_{i=1}^3q_i)\prod_{i=1}^3\frac{d\mathbf{q}_i}{2E_i(2\pi)^3}
\nonumber
\end{eqnarray}
with $E_1=\sqrt{m_\pi^2+\mathbf{q}_1^2}$,
$E_2=\sqrt{m_\pi^2+\mathbf{q}_2^2}$ and
$E_3=\sqrt{m_{\eta^{(\prime)}}^2+\mathbf{q}_3^2}$. After integrating
the delta function over the solid-angle elements $d\Omega_1$ and
$d\Omega_2$ and treating the X(1835) as an unpolarized particle, the
partial decay width can be expressed as a two dimensional integral
\begin{eqnarray}
\Gamma(X(1835) \to \eta^{(\prime)}\pi^+\pi^-) & = &
\frac{1}{64\pi^3m_X}\int|M^2|dE_1dE_2\label{3Width2Var}
\end{eqnarray}

The matrix elements are calculated by evaluating the loop integral.
For example, the matrix element $M_{(A)}$ for the corresponding
diagram (A) in Fig.~\ref{fig:strongdecayp} is
\begin{eqnarray}
iM_{(A)} & = &
-g_{eff}\frac{2}{(2m)^3}\int\frac{d^4k}{(2\pi)^4}\tilde{\Phi}((k-\frac{p}{2}))\nonumber\\
& & \times \frac{{\rm
Tr}\{(k\hspace{-0.2cm}\slash+m)\gamma_5[(k\hspace{-0.2cm}\slash-p\hspace{-0.2cm}\slash)+m]q\hspace{-0.2cm}\slash_3\gamma_5[(k\hspace{-0.2cm}\slash-q\hspace{-0.2cm}\slash_1-q\hspace{-0.2cm}\slash_2)+m]q\hspace{-0.2cm}\slash_2\gamma_5[(k\hspace{-0.2cm}\slash-q\hspace{-0.2cm}\slash_1)+m]q\hspace{-0.2cm}\slash_1\gamma_5\}}{(k^2-m^2)[(k-p)^2-m^2][(k-q_1-q_2)^2-m^2][(k-q_1)^2-m^2]}
\nonumber
\end{eqnarray}
where $g_{eff} =
g_{_X}g_{N\bar{N}\pi}^2g_{N\bar{N}\eta^{(\prime)}}$. After
performing the trace calculation, the matrix element can be
decomposed in terms of the tensor structure
\begin{eqnarray}
iM_{(A)} & = & -g_{eff}\frac{2}{(2m)^3}[4\alpha_0 D_0+4\alpha_\mu
D^\mu+4\alpha_{\mu\nu} D^{\mu\nu}+4\alpha_{\mu\nu\alpha}
D^{\mu\nu\alpha}] \nonumber
\end{eqnarray}
where $\alpha$'s are functions of the external momenta and $D$'s are
the loop integrals. Their explicit forms are given in
Appendix~\ref{App:LI}.

The results for the decay widths of $\Gamma(X \to \eta^{(\prime)}
\pi^+\pi^-)$ in the energy region $\Lambda_X = 2.0 - 3.0~$GeV are
\begin{eqnarray}
\Gamma(X \to \eta^\prime \pi^+\pi^-) = 0.580 - 1.273~{\rm MeV};
\;\;\;\; \Gamma^P(X \to \eta^\prime \pi^+\pi^-) = 0.335 -
0.400~{\rm MeV} \nonumber\\
\Gamma(X \to \eta \pi^+\pi^-) = 6.522 - 13.29~{\rm MeV}; \;\;\;\;\;
\Gamma^P(X \to \eta \pi^+\pi^-) = 1.550 - 1.926~{\rm MeV} \nonumber
\end{eqnarray}
where the upper index $P$ means that the results are from the pure
pseudoscalar processes illustrated in Fig.~\ref{fig:strongdecayp}.
The above decay widths increase with increase in $\Lambda_X$. To
obtain the above results, the coupling constant $g_{_{X}}$
calculated before and the coupling constants given above were used.
Using the central value of the total width
$\Gamma(X(1835))=67.7~$MeV~\cite{Ablikim:2005um}, the branching
ratios turn out to be
\begin{eqnarray}
{\rm BR}(X \to \eta^\prime \pi^+\pi^-) & \simeq &
8.57 \times 10^{-3} - 1.88 \times 10^{-2}  \nonumber\\
{\rm BR}(X \to \eta \pi^+\pi^-) & \simeq & 9.63 \times 10^{-2} -
1.96\times 10^{-1} \nonumber
\end{eqnarray}
Using the result ${\rm BR}(J/\psi\to\gamma X) \sim (0.5-2)\times
10^{-3}$~\cite{S.Jin}, the following product for branching fraction
is obtained
\begin{eqnarray}
{\rm BR}(J/\psi\to\gamma X){\rm BR}(X \to \eta^\prime \pi^+\pi^-)
\simeq (0.428 - 1.714)\times 10^{-5} - (0.94 - 3.76)\times 10^{-5}
\nonumber
\end{eqnarray}
which is much smaller than the observed data. The uncertainties in
the parentheses are from the uncertainty of ${\rm
BR}(J/\psi\to\gamma X)$. The large uncertainty comes from the
measurement of ${\rm BR}(J/\psi\to\gamma X)$. In addition, the
product of branching fraction ${\rm BR}(J/\psi\to\gamma X){\rm BR}(X
\to \eta \pi^+\pi^-)$ yields
\begin{eqnarray}
{\rm BR}(J/\psi\to\gamma X){\rm BR}(X \to \eta \pi^+\pi^-) \simeq
(0.418 - 1.926)\times 10^{-4} - (0.963 - 3.852)\times 10^{-4}
\nonumber
\end{eqnarray}
where the uncertainties in the parentheses are also from the
uncertainty of ${\rm BR}(J/\psi\to\gamma X)$.

Our calculation shows that the strong decay width $\Gamma(X(1835)
\to \eta^\prime \pi^+\pi^-)$ based on the baryonium assumption in
the energy scale $2.0{\rm ~GeV} \leq \Lambda_X \leq 3.0{\rm ~GeV}$
is much smaller than the data which leads to the conclusion that the
X(1835) may not be a baryonium. In addition, we have also predicted
the strong decay width of $\Gamma(X \to \eta \pi^+\pi^-)$ should be
larger than that of $\Gamma(X(1835) \to \eta^\prime \pi^+\pi^-)$ if
the X(1835) is a baryonium due to the large phase space and coupling
constant $g_{N\bar{N}\eta}$.

\subsection{Radiative decay of $X(1835) \to 2\gamma$.}

The X(1835) state can decay into two photons. Since the X(1835)
state is a pseudoscalar state the radiative decay is an anomalous
process. The matrix element can be written as
\begin{eqnarray}
iM^{\mu\nu}(X(1835) \to 2\gamma) & = & \alpha_{\rm
em}\epsilon_{\mu\nu\alpha\beta}p_\alpha q_\beta G_{X\gamma\gamma}
\nonumber
\end{eqnarray}
where $q$ and $p$ are the momenta of the two final photons. Using
the above expression the decay width is given by
\begin{eqnarray}
\Gamma(X(1835) \to 2\gamma) & = & \frac{1}{8\pi
m_X}|M|^2\frac{|\vec{p}_{cm}|}{m_X} = \frac{1}{32\pi}\alpha_{\rm
em}^2m_X^3 G_{X\gamma\gamma}^2 \nonumber
\end{eqnarray}
where $|\vec{p}_{\rm cm}| = m_X/2$ is the three-momentum of the
decay products.

\begin{figure}[htbp]
\begin{center}
\includegraphics[scale=0.7]{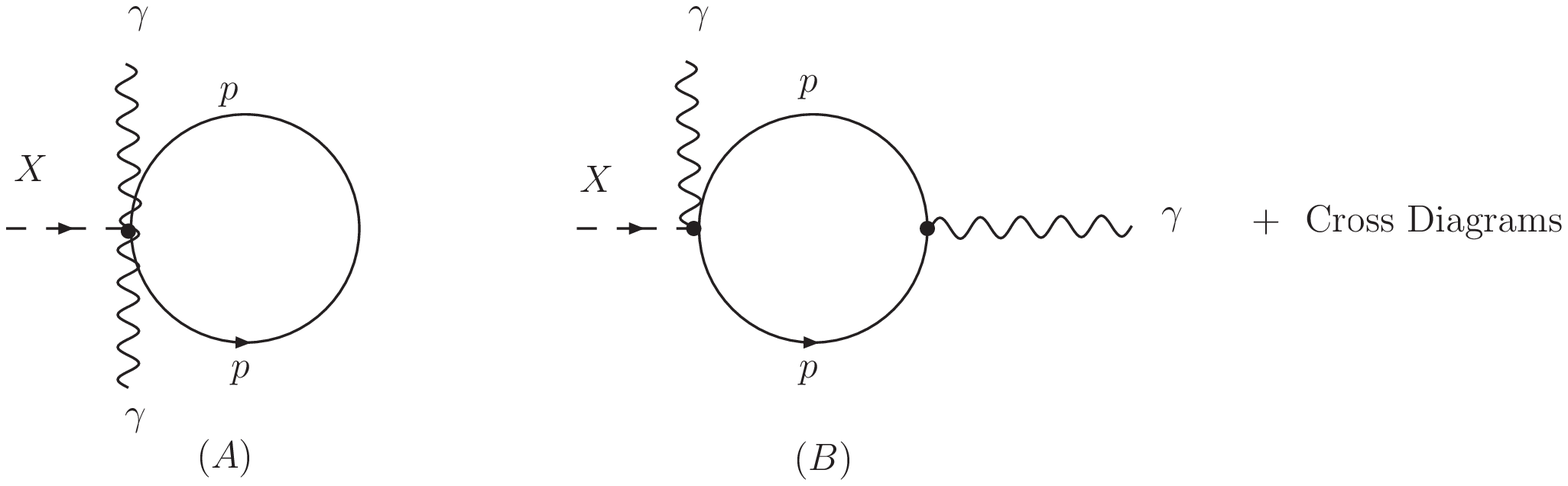}
\includegraphics[scale=0.7]{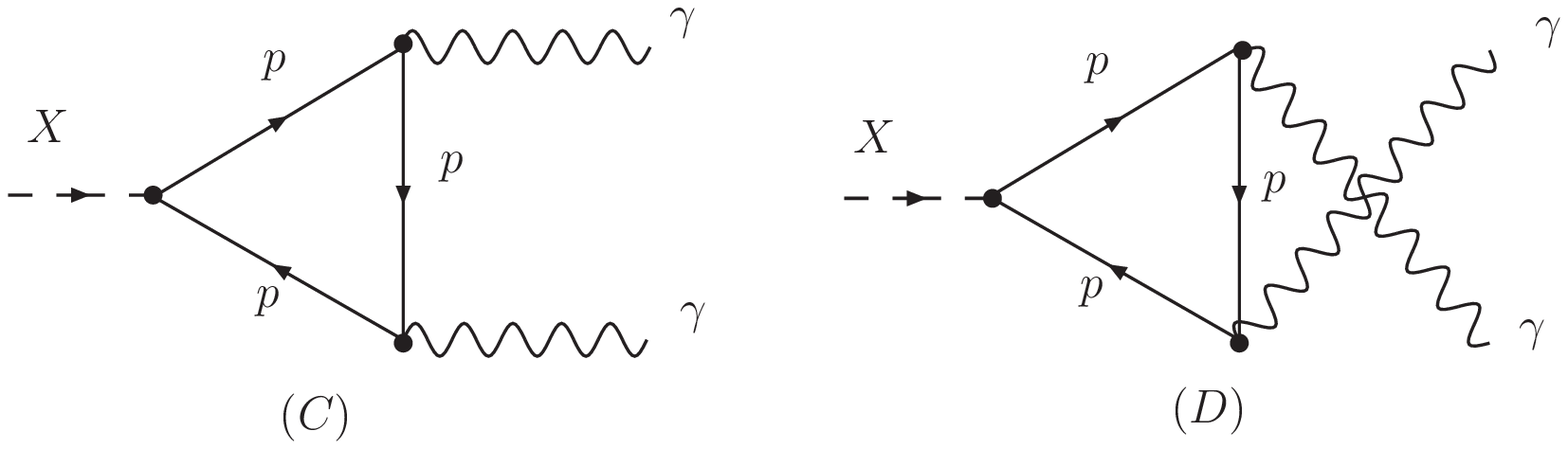}
\end{center}
\caption[Diagrams that  contributing to the radiative decay of $X(1835) \to 2\gamma$.]{%
Diagrams contributing to the radiative decay $X(1835) \to 2\gamma$.
} \label{fig:em}
\end{figure}

In our present model, the decay $X(1835) \to 2\gamma$ happens via
the process given by the diagrams in Fig.~\ref{fig:em}. Diagrams
$(A)$, $(B)$ and their corresponding cross diagrams arise from the
gauge of the nonlocal interaction (\ref{EMXNN}). Diagram $(A)$ and
its cross one are from quadratic terms of $A_\mu$ in the expansion
of Eq.~(\ref{EMXNN}) while diagram $(B)$ and its cross one are from
the linear terms of $A_\mu$ and the gauge of the proton free
Lagrangian (\ref{GaugeProton}). Diagrams $(C)$ and $(D)$ arise from
Lagrangian given by (\ref{GaugeProton}).

From our analysis neither diagram $(A)$ nor diagram $(B)$
contributes to the total matrix element due to the vanishing of the
trace of gamma matrices. So we need to calculate only the diagrams
$(C)$ and $(D)$ which are the same as that calculated in the
triangle anomaly problem. Since the discovery of the triangle
anomaly~\cite{Bell:1969ts,Adler:1969gk}, the calculation of these
diagrams have been discussed widely in the literature. We had
discussed the ambiguities in the calculations induced by
regularization, Dirac trace, and momentum shifts~\cite{Ma:2005md}.
From our calculation
\begin{eqnarray}
G_{X\gamma\gamma}^{\rm NL} & = &
g_{_X}\frac{2m_p}{\pi\Lambda_X^2}\int_0^\infty
d\alpha_1d\alpha_2d\alpha_3\frac{1}{(1+\alpha_1+\alpha_2+\alpha_3)^2}\exp\{{\rm
arg}_{\rm em}/\Lambda_X^2\} \nonumber
\end{eqnarray}
where the upper index ${\rm NL}$ corresponds to the Nonlocal case
and
\begin{eqnarray}
{\rm arg}_{\rm em} & = &
-\frac{1}{1+\sum_{i=1}^3\alpha_i}(\frac{1}{2}+\alpha_2)(\frac{1}{2}+\alpha_2+\alpha_3)m_X^2+(\frac{1}{4}+\alpha_2)m_X^2-\sum_{i=1}^3\alpha_im_p^2
\nonumber
\end{eqnarray}

Using the values of the parameters we present the numerical results
now. For the effective coupling $G_{X\gamma\gamma}^{\rm NL}$, and
for the scale parameter in the range $2.0{\rm ~GeV} \leq \Lambda_X
\leq 3.0{\rm ~GeV}$, we get the result
\begin{eqnarray}
G_{X\gamma\gamma}^{\rm NL} & = & 0.6813{\rm ~GeV}^{-1} - 0.3804{\rm
~GeV}^{-1} 
 \nonumber
\end{eqnarray}
and the corresponding electromagnetic decay width
\begin{eqnarray}
\Gamma_{X\gamma\gamma}^{\rm NL} & = & 1.516{\rm ~KeV} - 0.4726{\rm
~KeV} 
 \nonumber
\end{eqnarray}
Both $G_{X\gamma\gamma}^{\rm NL}$ and $\Gamma_{X\gamma\gamma}^{\rm
NL}$ decrease with increase in $\Lambda_X$.

The radiative decay has been investigated in Ref.~\cite{Li:2005vd}
treating the X(1835) as a glueball. The result obtained for the
decay width $\Gamma_{X\gamma\gamma} = 1.1(0.31-1.1)~$KeV agrees with
our result.

\section{discussions and conclusions}

\label{DisCon}

In this work, the strong decays of $X(1835) \to
\eta^{(\prime)}\pi^+\pi^-$ and electromagnetic decay of $X(1835) \to
2\gamma$ have been calculated using the effective Lagrangian method.
In our work we have treated the X(1835) as a baryonium. To fix the
only free parameter $\Lambda_X$ we postulated that the coupling
constant $g_{_X}$ has to be stable against $\Lambda_X$. With this
assumption, we varied $\Lambda_X$ from $2.0~$GeV to $3.0~$GeV. In
the above region the strong decay width of $X(1835) \to
\eta^\prime\pi^+\pi^-$ is much smaller than the observed data but
our prediction of the electromagnetic decay width of $X(1835) \to
2\gamma$ is in agreement with the result where X(1835) decays
through glueball. In addition, we have also calculated the strong
decay width $\Gamma(X(1835) \to \eta \pi^+\pi^-)$ explicitly. The
calculated width is much larger than the partial width of
$\Gamma(X(1835) \to \eta^\prime \pi^+\pi^-)$ which is consistent
with the direct analysis of the phase space and the coupling
constant.

In the baryonium picture, other decay modes of X(1835) can also be
calculated. Using the isospin relation we get
\begin{eqnarray}
\Gamma(X(1835) \to \eta^\prime\pi^0\pi^0) & = & \frac{1}{2}
\Gamma(X(1835) \to
\eta^\prime\pi^+\pi^-) = 0.290 - 0.637~\mbox{MeV} ; \nonumber\\
\Gamma(X(1835) \to \eta\pi^0\pi^0) & = & \frac{1}{2} \Gamma(X(1835)
\to \eta\pi^+\pi^-) = 3.261 - 6.645~\mbox{MeV} \nonumber
\end{eqnarray}
The other three-pseudoscalar strong decay channels are either
isospin symmetry violating processes ($\pi^+\pi^-\pi^0$ and
$3\pi^0$) or OZI rule suppressed (with Kaon meson in the final
state). The four strong decay channels discussed above are dominant
among all the three-pseudoscalar channels. We have listed the
effective coupling constant $g_{_{X}}$ and their decay widths in the
region $~2.0{\rm GeV}\leq \Lambda_X\leq 3.0{~\rm GeV}$ in
Table.~\ref{table:lambdadep}.
\begin{table}
\begin{center}
\begin{tabular}{|r|r|r|r|r|r|r|r|r|r|}
\hline
$ g_{_{X}}~~~~$ & $\Gamma_{X \to \eta^\prime \pi^+\pi^-}$ & $\Gamma_{X \to \eta \pi^+\pi^-}$ & $\Gamma_{X \to \eta^\prime \pi^0\pi^0}$ & $\Gamma_{X \to \eta \pi^0\pi^0}$ & $\Gamma_{X \to 2\gamma}$~~~~ \\
\hline  2.55 $-$ 2.37 & 0.580 $-$ 1.273 & 6.522 $-$ 13.29 & 0.290 $-$ 0.637 & 3.261 $-$ 6.645 & 1.516 $-$ 0.4726\\
\hline
\end{tabular}
\caption{\label{table:lambdadep} The $\Lambda_X$ dependence of
effective coupling constant and decay widths in the region $2.0{~\rm
GeV} \leq \Lambda_X \leq 3.0{~\rm GeV}$(The strong decay width is
expressed in unit of MeV while the electromagnetic decay width is
expressed in unit of KeV).}
\end{center}
\end{table}

It should be noticed that in principal, the finite width effect
should be included by introducing the Breit-Wigner distribution
function. However, this will suppress our results and our final
conclusion will not be changed. Moreover, there are also
uncertainties from the sigma meson mass and width. Here, we applied
the results yielded by unitarizing the $\pi\pi$ and $\pi K$
scattering amplitudes.

To conclude, we have studied the three-pseudoscalar meson and
two-photon decays of X(1835). The strong decay width $\Gamma(X(1835)
\to \eta^\prime \pi^+\pi^-)$ is smaller than the experimental data
while the two-photon decay width agrees with the result where
X(1835) was assumed to decay via the glueball assumption. From our
results X(1835) cannot be treated as a baryonium, at least in the
framework of the composite model as applied in this paper. We have
obtained other dominant three-pseudoscalar meson decay channels from
the isospin relations. To confirm the structure of $X(1835)$ further
theoretical analysis is necessary.

\appendix

\section{Decomposition of one loop integral.}

\label{App:LI}

For the one loop integral of diagram (A) of
Fig.~\ref{fig:strongdecayp}, after performing the trace calculation
we get the following decomposition
\begin{eqnarray}
iM_{(A)} & = &
-g_{eff}\frac{2}{(2m)^3}\int\frac{d^4k}{(2\pi)^4}\tilde{\Phi}((k-\frac{p}{2}))\nonumber\\
& & \times \frac{{\rm
Tr}\{(k\hspace{-0.2cm}\slash+m)\gamma_5[(k\hspace{-0.2cm}\slash-p\hspace{-0.2cm}\slash)+m]q\hspace{-0.2cm}\slash_3\gamma_5[(k\hspace{-0.2cm}\slash-q\hspace{-0.2cm}\slash_1-q\hspace{-0.2cm}\slash_2)+m]q\hspace{-0.2cm}\slash_2\gamma_5[(k\hspace{-0.2cm}\slash-q\hspace{-0.2cm}\slash_1)+m]q\hspace{-0.2cm}\slash_1\gamma_5\}}{(k^2-m^2)[(k-p)^2-m^2][(k-q_1-q_2)^2-m^2][(k-q_1)^2-m^2]}\nonumber\\
& = & -g_{eff}\frac{2}{(2m)^3}[4\alpha_0 D_0+4\alpha_\mu
D^\mu+4\alpha_{\mu\nu} D^{\mu\nu}+4\alpha_{\mu\nu\alpha}
D^{\mu\nu\alpha}] \nonumber
\end{eqnarray}
where
\begin{eqnarray}
\alpha^0 & = & m^3\bigg[p\cdot q_3q_1\cdot q_2-p\cdot q_2q_1\cdot q_3-2q_1\cdot q_2q_1\cdot q_3-q_1\cdot q_3 q_2^2+p\cdot q_1q_2\cdot q_3+2q_1^2q_2\cdot q_3\bigg]\nonumber\\
& & + m\bigg[-p\cdot q_3 q_1\cdot q_2 q_1^2-p\cdot q_2q_1\cdot q_3q_1^2-p\cdot q_3q_1^2q_2^2+p\cdot q_1q_2\cdot q_3q_1^2\bigg] \nonumber\\
\alpha^\mu & = & m^3\bigg[q_1\cdot q_2 q_3^\mu -2q_2\cdot q_3q_1^\mu\bigg]\nonumber\\
& & + m \bigg[-2p\cdot q_2 q_1^2q_3^\mu + 4p\cdot q_3q_1^2q_2^\mu+4p\cdot q_1 q_1\cdot q_2 q_3^\mu-2p\cdot q_3q_1\cdot q_2q_1^\mu+2p\cdot q_2q_1\cdot q_3q_1^\mu\nonumber\\
& & \;\;\;\;\;\;\;\;\; - 4q_1\cdot q_2q_1\cdot q_3 p^\mu+2p\cdot q_1q_2^2q_3^\mu-2q_1\cdot q_3q_2^2p^\mu-2p\cdot q_1 q_2\cdot q_3q_1^\mu+2q_2\cdot q_3q_1^2p^\mu\bigg] \nonumber\\
\alpha^{\mu\nu} & = & m\bigg[-4p\cdot q_1q_2^\mu q_3^\nu+4p\cdot q_2q_1^\mu q_3^\nu-4p\cdot q_3q_1^\mu q_2^\nu+4q_1\cdot q_3 p^\mu q_2^\nu -4q_2\cdot q_3p^\mu q_1^\nu\bigg] \nonumber\\
& & + m\bigg[3p\cdot q_3q_1\cdot q_2-3p\cdot q_2q_1\cdot q_3+ 2q_1\cdot q_2q_1\cdot q_3+q_1\cdot q_3 q_2^2+3p\cdot q_1q_2\cdot q_3-2q_1^2q_2\cdot q_3\bigg]g^{\mu\nu}\nonumber\\
\alpha^{\mu\nu\alpha} & = & 2m\bigg[q_2\cdot q_3q_1^\alpha-q_1\cdot
q_2 q_3^\alpha\bigg]g^{\mu\nu} \nonumber
\end{eqnarray}
and
\begin{eqnarray}
D_{\{0;\mu;\mu\nu;\mu\nu\alpha\}} & = &
\int\frac{d^4k}{(2\pi)^4}\tilde{\Phi}((k-\frac{p}{2}))
\frac{\{1;k_\mu;k_\mu k_\nu;k_\mu k_\nu
k_\alpha\}}{(k^2-m^2)[(k-p)^2-m^2][(k-q_1-q_2)^2-m^2][(k-q_1)^2-m^2]}
\nonumber
\end{eqnarray}

It is to be noted that due to the relation
\begin{eqnarray}
k\cdot p & = &
-\frac{1}{2}\bigg\{[(k-p)^2-m^2]-(k^2-m^2)-p^2\bigg\} \nonumber\\
k\cdot q_1 & = &
-\frac{1}{2}\bigg\{[(k-q_1)^2-m^2]-(k^2-m^2)-q_1^2\bigg\} \nonumber\\
k\cdot q_3 & = &
-\frac{1}{2}\bigg\{[(k-p)^2-m^2]-[(k-q_1-q_2)^2-m^2]+(q_1+q_2)^2\bigg\}
\nonumber
\end{eqnarray}
the above vector, two- and three- rank four-point integrals can be
expressed in terms of scalar four-point and three-point integrals
\begin{eqnarray}
\alpha_{\mu} D^{\mu}  & = & \beta^{V}_0D_0 + \beta_{234}^{V}C_{234;0} + \beta_{134}^{V}C_{134;0} + \beta_{124}^{V}C_{124;0} + \beta_{123}^{V}C_{123;0} \nonumber\\
\alpha_{\mu\nu} D^{\mu\nu}  & = & m^3\beta_0^{T1}D_0 +m\beta_0^{T1}C_{234;0} + m \bigg[\beta_{123}^{T1;\mu} C_{123;\mu}+\beta_{124}^{T1;\mu} C_{124;\mu}+\beta_{134}^{T1;\mu} C_{134;\mu}+\beta_{234}^{T1;\mu} C_{234;\mu}\bigg]\nonumber\\
& & \;\;\;\;\;\;\;\;\;\;\;\;\;\;\;\;\;\;\;\;\;\;\;\;\;\;\;\;\;\;\;\;\;\;\;\;\;\;\;  + m \bigg[ \beta_0^{T2} D_{0} + \beta_{123}^{T2} C_{123;0} + \beta_{124}^{T2} C_{124;0} + \beta_{234}^{T2} C_{234;0}\bigg] \nonumber\\
\alpha_{\mu\nu\alpha} D^{\mu\nu\alpha} & = & 2m\bigg[q_2\cdot
q_3q_1^\mu-q_1\cdot q_2q_3^\mu\bigg]C_{234;\mu} - m^3\bigg[q_2\cdot
q_3[C_{123;0}-C_{234;0}]-q_1\cdot
q_2[C_{134;0}-C_{124;0}]\nonumber\\
& &
\;\;\;\;\;\;\;\;\;\;\;\;\;\;\;\;\;\;\;\;\;\;\;\;\;\;\;\;\;\;\;\;\;\;\;\;\;\;\;\;\;\;\;\;\;\;\;\;\;\;\;\;\;\;\;\;\;\;\;
- \{q_2\cdot q_3q_1^2+q_1\cdot q_2[(q_1+q_2)^2-p^2]\}D_0\bigg]
\nonumber
\end{eqnarray}
with
\begin{eqnarray}
\beta^{V}_0 & = & - \frac{1}{2} \bigg\{m^3 \bigg[ q_1 \cdot q_2[(q_1+q_2)^2 - q^2] + 2q_2\cdot q_3 q_1^2 \bigg] \nonumber\\
& & \;\;\;\;\;\;\; + m \bigg[ -2 q\cdot q_2 q_1^2[(q_1 +q_2)^2 - q^2] + 4 q\cdot q_3 q_1^2 [q_1^2 -(q_1 + q_2)^2] \nonumber\\
& & \;\;\;\;\;\;\;\;\;\;\;\;\;\;\; + 4 q\cdot q_1 q_1\cdot q_2[(q_1 + q_2)^2-q^2] + 2 q\cdot q_3 q_1\cdot q_2 q_1^2 - 2q\cdot q_2 q_1\cdot q_3 q_1^2 \nonumber\\
& & \;\;\;\;\;\;\;\;\;\;\;\;\;\;\; + 4 q_1 \cdot q_2 q_1 \cdot q_3 q^2 + 2 q\cdot q_1 q_2^2 [(q_1 + q_2)^2 -q^2] + 2 q_1 \cdot q_3 q_2^2 q^2 + 2 q\cdot q_1 q_2 \cdot q_3 q_1^2 \nonumber\\
& & \;\;\;\;\;\;\;\;\;\;\;\;\;\;\; - 2 q_2\cdot q_3 q_1^2 q^2\bigg]\bigg\} \nonumber\\
\beta_{234}^{V} & = & - \frac{1}{2} \bigg\{2m^3 q_2\cdot q_3 + m \bigg[ 2 q\cdot q_3 q_1\cdot q_2 - 2q\cdot q_2 q_1\cdot q_3 + 4 q_1 \cdot q_2 q_1 \cdot q_3 + 2 q_1 \cdot q_3 q_2^2 + 2 q\cdot q_1 q_2 \cdot q_3 \nonumber\\
& & \;\;\;\;\;\;\;\;\;\;\;\;\;\;\;\;\;\;\;\;\;\;\;\;\;\;\;\;\;\;\; - 2 q_2\cdot q_3 q_1^2 \bigg]\bigg\}\nonumber\\
\beta_{134}^{V} & = &  - \frac{1}{2} \bigg\{m^3 q_1 \cdot q_2 + m \bigg[ -2 q\cdot q_2 q_1^2 + 4 q\cdot q_1 q_1\cdot q_2 - 4 q_1\cdot q_2 q_1\cdot q_3 + 2 q\cdot q_1 q_2^2 - 2 q_1\cdot q_3 q_2^2 \nonumber\\
& & \;\;\;\;\;\;\;\;\;\;\;\;\;\;\;\;\;\;\;\;\;\;\;\;\;\;\;\;\;\; + 2 q_2\cdot q_3 q_1^2 \bigg]\bigg\} \nonumber\\
\beta_{124}^{V} & = & - \frac{1}{2} \bigg\{ - m^3 q_1 \cdot q_2 + m \bigg[ 2 q\cdot q_2 q_1^2 + 4 q\cdot q_3 q_1^2 - 4 q\cdot q_1 q_1\cdot q_2 - 2 q \cdot q_1 q_2^2 \bigg]\bigg\} \nonumber\\
\beta_{123}^{V} & = & - \frac{1}{2} \bigg\{ - 2m^3 q_2 \cdot q_3 + m \bigg[ -4 q\cdot q_3 q_1^2 - 2 q\cdot q_3q_1\cdot q_2 + 2q\cdot q_2 q_1\cdot q_3 - 2 q\cdot q_1 q_2 \cdot q_3 \bigg]\bigg\} \nonumber\\
\beta_0^{T1} & = & 3p\cdot q_3q_1\cdot q_2-3p\cdot q_2q_1\cdot q_3+ 2q_1\cdot q_2q_1\cdot q_3+q_1\cdot q_3 q_2^2+3p\cdot q_1q_2\cdot q_3-2q_1^2q_2\cdot q_3 \nonumber\\
\beta_{123;\mu}^{T1} & = & 2p\cdot q_3q_2^\mu \nonumber\\
\beta_{124;\mu}^{T1} & = & 2p\cdot q_2q_1^\mu - 2p\cdot q_1q_2^\mu  \nonumber\\
\beta_{134;\mu}^{T1} & = & 2p\cdot q_1q_2^\mu-2p\cdot q_2q_1^\mu-2q_1\cdot q_3q_2^\mu+2q_2\cdot q_3q_1^\mu \nonumber\\
\beta_{234;\mu}^{T1} & = & 2q_1\cdot q_3q_2^\mu-2p\cdot q_3
q_2^\mu-2q_2\cdot q_3q_1^\mu  \nonumber\\
\beta_0^{T2} & = & - \frac{1}{2}\bigg\{ 2 q\cdot q_1[(q_1+q_2)^2 - q^2][-(q_1+q_2)^2 + q_1^2] + 2 q\cdot q_2 q_1^2[(q_1+q_2)^2 - q^2]\nonumber\\
& & - 2 q\cdot q_3 q_1^2[-(q_1+q_2)^2 + q_1^2] + 2 q_1\cdot q_3 q^2 [-(q_1+q_2)^2 + q_1^2] + 2 q_2\cdot q_3 q^2 q_1^2\bigg\}  \nonumber\\
\beta_{123}^{T2} & = & - \frac{1}{2}\bigg\{ - 2 q\cdot q_1[(q_1+q_2)^2 - q^2] - 2 q\cdot q_2 [(q_1+q_2)^2 - q^2]\nonumber\\
& & \;\;\;\;\;\;\;\;\; + 2 q\cdot q_3 q_1^2 - 2 q_1\cdot q_3 q^2 - 2 q_2\cdot q_3 q^2 \bigg\}  \nonumber\\
\beta_{124}^{T2} & = & - \frac{1}{2}\bigg\{ 2 q\cdot q_1[(q_1+q_2)^2 - q^2] - 2 q\cdot q_3 q_1^2 + 2 q_1\cdot q_3 q^2 \bigg\}  \nonumber\\
\beta_{234}^{T2} & = & - \frac{1}{2}\bigg\{ 2 q\cdot q_2
[(q_1+q_2)^2 - q^2] + 2 q_2\cdot q_3 q^2 \bigg\} \nonumber
\end{eqnarray}
and
\begin{eqnarray}
C_{123;\{0;\mu\}} & = & \int\frac{d^4k}{(2\pi)^4}\tilde{\Phi}((k-\frac{p}{2})) \frac{\{1;k_\mu\}}{(k^2-m^2)[(k-p)^2-m^2][(k-q_1-q_2)^2-m^2]} \nonumber\\
C_{124;\{0;\mu\}} & = & \int\frac{d^4k}{(2\pi)^4}\tilde{\Phi}((k-\frac{p}{2})) \frac{\{1;k_\mu\}}{(k^2-m^2)[(k-p)^2-m^2][(k-q_1)^2-m^2]} \nonumber\\
C_{234;\{0;\mu\}} & = & \int\frac{d^4k}{(2\pi)^4}\tilde{\Phi}((k-\frac{p}{2})) \frac{\{1;k_\mu\}}{[(k-p)^2-m^2][(k-q_1-q_2)^2-m^2][(k-q_1)^2-m^2]} \nonumber\\
C_{134;\{0;\mu\}} & = &
\int\frac{d^4k}{(2\pi)^4}\tilde{\Phi}((k-\frac{p}{2}))
\frac{\{1;k_\mu\}}{(k^2-m^2)[(k-q_1-q_2)^2-m^2][(k-q_1)^2-m^2]}
\nonumber
\end{eqnarray}

Using the above, the matrix element $iM_{(A)}$ can be expressed in
terms of the scalar, and vector $C$ and $D$ functions. For the
scalar, vector $C$ and $D$ functions one can evaluate the momentum
integral explicitly and yield the following results.
\begin{eqnarray}
D_{\{0;\mu\}} & = &
\frac{i}{16\pi^2}\frac{1}{\Lambda_X^4}\int_0^\infty
d\alpha_1d\alpha_2d\alpha_3d\alpha_4\frac{1}{(1+\tilde{\alpha}_{4})^2}\{1;\frac{1}{1+\tilde{\alpha}_{4}}P_{D;\mu}\}\exp\{{\rm
arg}_D/\Lambda_X^2\} \nonumber\\
C_{123;\{0;\mu\}} & = &
-\frac{i}{16\pi^2}\frac{1}{\Lambda_X^2}\int_0^\infty
d\alpha_1d\alpha_2d\alpha_3\frac{1}{(1+\tilde{\alpha}_{3})^2}\{1;\frac{1}{1+\tilde{\alpha}_{3}}P_{C_{123};\mu}\}\exp\{{\rm
arg}_{C_{123}}/\Lambda_X^2\} \nonumber\\
C_{124;\{0;\mu\}} & = &
-\frac{i}{16\pi^2}\frac{1}{\Lambda_X^2}\int_0^\infty
d\alpha_1d\alpha_2d\alpha_3\frac{1}{(1+\tilde{\alpha}_{3})^2}\{1;\frac{1}{1+\tilde{\alpha}_{3}}P_{C_{124};\mu}\}\exp\{{\rm
arg}_{C_{124}}/\Lambda_X^2\} \nonumber\\
C_{234;\{0;\mu\}} & = &
-\frac{i}{16\pi^2}\frac{1}{\Lambda_X^2}\int_0^\infty
d\alpha_1d\alpha_2d\alpha_3\frac{1}{(1+\tilde{\alpha}_{3})^2}\{1;\frac{1}{1+\tilde{\alpha}_{3}}P_{C_{234};\mu}\}\exp\{{\rm
arg}_{C_{234}}/\Lambda_X^2\} \nonumber\\
C_{134;\{0;\mu\}} & =
&-\frac{i}{16\pi^2}\frac{1}{\Lambda_X^2}\int_0^\infty
d\alpha_1d\alpha_2d\alpha_3\frac{1}{(1+\tilde{\alpha}_{3})^2}\{1;\frac{1}{1+\tilde{\alpha}_{3}}P_{C_{134};\mu}\}\exp\{{\rm
arg}_{C_{134}}/\Lambda_X^2\} \nonumber
\end{eqnarray}
where
\begin{eqnarray}
\tilde{\alpha}_{n} & = &\sum_{i=1}^n\alpha_i \nonumber\\
P_{D;\mu} & = &
(\frac{1}{2}+\alpha_2)p_\mu+\alpha_3q_{1;\mu}+\alpha_4(q_1+q_2)_\mu \nonumber\\
P_{C_{123};\mu} & = & (\frac{1}{2}+\alpha_2)p_\mu+\alpha_3(q_1+q_2)_\mu \nonumber\\
P_{C_{124};\mu} & = & (\frac{1}{2}+\alpha_2)p_\mu+\alpha_3q_{1;\mu} \nonumber\\
P_{C_{234};\mu} & = & (\frac{1}{2}+\alpha_1)p_\mu+\alpha_2(q_1+q_2)_\mu+\alpha_3q_{1;\mu} \nonumber\\
P_{C_{134};\mu} & = & \frac{1}{2}p_\mu+\alpha_2(q_1+q_2)_\mu+\alpha_3q_{1;\mu} \nonumber\\
{\rm arg}_{D} & = &
-\frac{1}{1+\tilde{\alpha}_4}P_D^2+(\frac{1}{4}+\alpha_2)p^2+\alpha_3q_1^2+\alpha_4(q_1+q_2)^2-\tilde{\alpha}_4m^2 \nonumber\\
{\rm arg}_{C_{123}} & = &
-\frac{1}{1+\tilde{\alpha}_3}P_{C_{123}}^2+(\frac{1}{4}+\alpha_2)p^2+\alpha_3(q_1+q_2)^2-\tilde{\alpha}_3m^2 \nonumber\\
{\rm arg}_{C_{124}} & = &
-\frac{1}{1+\tilde{\alpha}_3}P_{C_{124}}^2+(\frac{1}{4}+\alpha_2)p^2+\alpha_3q_1^2-\tilde{\alpha}_3m^2 \nonumber\\
{\rm arg}_{C_{234}} & = &
-\frac{1}{1+\tilde{\alpha}_3}P_{C_{234}}^2+(\frac{1}{4}+\alpha_1)p^2+\alpha_2(q_1+q_2)^2+\alpha_3q_1^2-\tilde{\alpha}_3m^2 \nonumber\\
{\rm arg}_{C_{134}} & = &
-\frac{1}{1+\tilde{\alpha}_3}P_{C_{134}}^2+\frac{1}{4}p^2+\alpha_2(q_1+q_2)^2+\alpha_3q_1^2-\tilde{\alpha}_3m^2
\nonumber
\end{eqnarray}
\acknowledgments

\label{ACK}

I would like to thanks Profs.~Amand Faessler, Thomas Gutsche and
Yu-Peng Yan for valuable discussions we had with them. We also thank
Prof.~Yue-Liang Wu(ITP, CAS) for suggesting the problem. This work
was supported by International Graduiertenkolleg der DFG GRK683
"Hadronen im Vakuum, in Kernen und in Sternen".


\end{document}